\documentclass[%
 reprint,
showpacs,preprintnumbers,
 amsmath,amssymb,
 aps,pra
]{revtex4-1}

\usepackage{graphicx}
\usepackage{dcolumn}
\usepackage{bm}

\newcommand{\ave}[1]{\ensuremath{\langle #1 \rangle}}

\begin{document}

\title{Analyzing the Gaussian character of the spectral quantum state of light via quantum noise measurements}

\author{A. S. Coelho$^1$, F. A. S. Barbosa$^1$, K. N. Cassemiro$^2$, M. Martinelli$^1$, A. S. Villar$^2$, and P. Nussenzveig$^1$}

\email{villar@df.ufpe.br}

\affiliation{$^1$ Instituto de F\'\i sica, Universidade de S\~ao Paulo, P.O. Box 66318, 
05315-970 S\~ao Paulo, SP, Brazil \\ 
$^2$ Departamento de F\'\i{}sica, Universidade Federal de Pernambuco, 50670-901 Recife, PE, Brazil
}

\begin{abstract}
Gaussian quantum states hold special importance in the continuous variable (CV) 
regime. In quantum information science, the understanding and characterization 
of central resources such as entanglement may strongly rely on the knowledge 
of the Gaussian or non-Gaussian character of the quantum state. However, the 
quantum measurement associated with the spectral photocurrent of light modes 
consists of a mixture of quadrature observables. Within the framework of two 
recent papers [Phys. Rev. A \textbf{88}, 052113 (2013) and Phys. Rev. 
Lett. \textbf{111}, 200402 (2013)], we address here how the statistics of 
the spectral photocurrent relates to the character of the Wigner function 
describing those modes. We show that a Gaussian state can be misidentified 
as non-Gaussian and vice-versa, a conclusion that forces the adoption of 
tacit \textit{a priori} assumptions to perform quantum state reconstruction. 
We experimentally analyze the light beams generated by the optical parametric 
oscillator (OPO) operating above threshold to show that the data strongly 
supports the generation of Gaussian states of the field, validating the use 
of necessary and sufficient criteria to characterize entanglement in this system. 
\end{abstract}


\maketitle 

\section{Introduction}
\indent

Entanglement and non-Gaussian quantum states are prized resources in the field of 
quantum information with continuous variables (CV). While Gaussian states 
\textit{per se} allow the generation of highly entangled states and thus the 
realization of important quantum information 
protocols~\cite{furusawaTeleportNature04,pengEntSwapPRL04}, certain tasks 
such as quantum computation and entanglement distillation require non-Gaussian 
features, and thus the ability to correctly recognize the availability of those 
resources in actual physical systems~\cite{nonGaussianComputation,nonGaussianDistillation}. 

Light is perhaps the most important physical system for CV quantum information. 
Measurement of the quantum state of light is realized by the statistical analysis 
of photocurrent fluctuations, the \textit{quantum noise}~\cite{cavesAmplifiersPRA82,slusherFirstSQZPRL85,heidmannPRL87,levenson4modesqzPRL87,oupereirakimblePRL92,raymertomoPRL93,mlynekNature97}. 
The physical objects of interest are often field modes with well defined frequency, 
in which case the spectral noise power of photocurrent fluctuations is used to 
provide information about the field quantum state. 

However, owing to a lack of phase coherence between the quantum state and the spectral 
components of the photocurrent signal, current measurement techniques only provide 
a pure quantum measurement for a restricted class of quantum states possessing a strong 
degree of symmetry~\cite{ralphSinglePhotonSidebandsPRA08,prl2013,pra2013}. In most experiments, the quantum observable 
associated with the spectral photocurrent provides a mixed measurement of the two sideband 
modes and requires the adoption of \textit{a priori} assumptions to interpret the data 
in terms of moments of field quadrature operators. These limitations constitute a problem 
for the demonstration of quantum information protocols with truly general and unknown 
quantum states of spectral light modes. 

Yet, it is generally believed that Gaussian spectral photocurrent statistics can be 
taken as \textit{proof} of the Gaussian character of the quantum state. Conversely, the 
observation of non-Gaussian photocurrent statistics is accepted as strong evidence of 
non-Gaussian features of the field quantum state~\cite{nonGaussianDAuria}. Contrarily 
to this belief, we show that the mixed character of the photocurrent measurement operator 
fundamentally \textit{masks} the phase space statistics of field quadratures, implying 
that a Gaussian state may appear non-Gaussian and vice-versa. Such dubiety about the 
statistics obeyed by the CV quantum state is especially harmful to experimental quantum 
information, where non-Gaussian properties of field modes stand as a necessary resource 
to perform more powerful tasks: misidentifying those resources leads to incorrect or, 
at least, unreliable implementations. 

In this paper, we perform a formal analysis of the measurement operator moments associated 
with the spectral photocurrent to evaluate the extent to which the observed statistics 
can be interpreted in terms of quantum state features. We consider in Sec.~\ref{sec:gaussianity} 
the photocurrent fluctuations as an incoherent mixture of two independent quantum measurements 
to show, by exploiting higher-order moments, that a Gaussian photocurrent can be produced 
either by Gaussian states with \textit{spectral two-mode symmetry} or by very peculiar 
(and hence implausible in most experiments) non-Gaussian quantum states. In fact, our 
reasoning links the Gaussian character and the symmetry of the two-mode 
field as equivalent \textit{a priori} knowledge always needed in the usual experimental 
setup (but rarely mentioned) to reconstruct the spectral quantum state. We 
additionally point out that the observation of non-Gaussian spectral photocurrent can 
not be readily associated with a non-Gaussian quantum state, requiring further investigation 
and possibly the improvement of the quantum measurement technique of CV spectral modes to 
a new level of experimental rigor. 

We then employ our methods in Sec.~\ref{sec:experiment} to experimentally investigate 
the quantum state produced by the optical parametric oscillator (OPO) operating above 
the oscillation threshold, establishing with great confidence that the associated 
photocurrent is indeed Gaussian and indicates a Gaussian quantum state displaying modal 
symmetries in the spectral modes linked by the parametric process. This conclusion 
substantiates the use of the semi-classical treatment in the above-threshold OPO to 
effectively halve the number of relevant modes and apply necessary and sufficient 
criteria for multipartite entanglement directly to the spectral 
matrix~\cite{villarPRL05,science09,natphoton10}. It is our feeling that such an 
experimental demonstration fills a hole that went unrecognized for some time. 

Finally, to guarantee that the measurement operator of the spectral photocurrent 
yields the statistics of pure field quadrature observables, we offer our 
concluding remarks in Sec.~\ref{sec:conclusion}, where the use of phase coherent 
optical and electronic local electronic oscillators is proposed to realize 
a \textit{doubly phase-coherent detection}. Only in this improved scenario 
can the quantum noise achieve the status of a formal pure measurement operator 
in the CV regime of spectral modes of light.

\section{Gaussian photocurrent and properties of the spectral field modes quantum state}
\label{sec:gaussianity}

\textit{Spectral photocurrent operator} -- The quantum state of spectral (sideband) modes is measured by the \textit{photocurrent} observable~\cite{wolf}, described by the measurement operator $\hat I(t)$ and obtained by the beating of the optical local oscillator (LO) and the field modes of interest~\cite{yuenshapiroHD80}. 
Any Fourier component of this signal can be extracted by mixing it with a sinusoidal electronic reference -- the electronic local oscillator (eLO) -- at frequency $\Omega$, yielding the spectral photocurrent operator $\hat I_\Omega$ defined as
\begin{equation}
\label{eq:defIomega}
\hat I_\Omega(t) = {\textstyle\frac{1}{\sqrt2}}\int_t^{t+T} \mathrm{e}^{i\Omega t'}\hat I(t') dt'. 
\end{equation}
The operator $\hat I_\Omega(t)$ represents the (complex) amplitude of the photocurrent beatnote signal measured at time $t$. 
The integration limits are determined by the spectral resolution $\Delta\Omega=2\pi/T$ of the electronic downmixing process and define the spectral width of measured upper $\omega_0+\Omega$ and lower $\omega_0-\Omega$ sideband modes. 
\begin{figure}[htbp]
\centerline{\includegraphics[width=0.7\columnwidth]{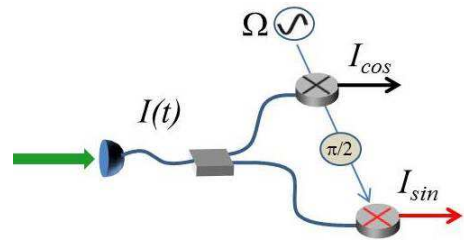}}
\caption{Scheme to measure electronic quadrature
components of each photocurrent signal. The photocurrent is mixed
with two electronic references in quadrature.}
\label{fig:detsetup}
\end{figure}
The spectral photocurrent operator $\hat I_\Omega(t)$ gathers in one quantity the two observables associated with the photocurrent cosine $\hat I_\mathrm{cos}$ and sine $\hat I_\mathrm{sin}$ components, given by 
\begin{equation}
\hat I_\Omega = {\textstyle\frac{1}{\sqrt2}}\,(\hat I_{\cos{}} + i \hat I_{\sin{}}).
\end{equation}
We may interpret $\hat I_\mathrm{cos}(t)$ and $\hat I_\mathrm{sin}(t)$ as two independent single quantum measurements taken at time $t$~\cite{pra2013}. Fig. \ref{fig:detsetup} illustrates the measurement apparatus. These observables access the quadrature operators of optical sideband modes at frequencies $\omega_0\pm\Omega$, where $\omega_0$ is the LO optical frequency, and bandwidth $\Delta\Omega$. These longitudinal modes involve the annihilation operators $\hat a_{\pm\Omega} = (\hat p_{\pm\Omega} + i \hat q_{\pm\Omega})/2$, where $\hat p_{\pm\Omega}$ and $\hat q_{\pm\Omega}$ are respectively the field amplitude and phase quadrature observables satisfying the canonical commutation relation $[\hat p_{\Omega},\hat q_{\Omega'}]=2i\delta(\Omega-\Omega')$. The interpretation of the photocurrent observables $\hat I_\mathrm{cos}$ and $\hat I_\mathrm{sin}$ in terms of the spectral quantum state is easily described after a change of modal basis by the symmetric ($\mathcal{S}$) and anti-symmetric ($\mathcal{A}$) combinations of sideband modes, defined as 
\begin{align}
\hat a_{s} &= {\textstyle\frac{1}{\sqrt2}}(\hat a_{\omega_0+\Omega}+\hat a_{\omega_0-\Omega}),\nonumber\\
\hat a_{a} &= {\textstyle\frac{1}{\sqrt2}}(\hat a_{\omega_0+\Omega}-\hat a_{\omega_0-\Omega}).
\label{eq:aadagdef}
\end{align}
The exact connection between the quadrature observables of these modes, defined by the expressions $\hat a_s = \hat p_s+ i \hat q_s$ and $\hat a_a = \hat p_a+ i \hat q_a$, and the photocurrent observables $\hat I_\mathrm{cos}$ and $\hat I_\mathrm{sin}$ depends on the measurement technique in use. 
In homodyne detection (HD), the cosine observable is a pure quantum measurement of the form $\hat I_{\cos{}} = L_{\cos{}}(\hat p_{s},\hat q_{s})$, where $L_{\cos{}}$ represents a linear combination; the sine observable has the same general form, but depends only on quadratures of the $\mathcal{A}$ mode. In resonator detection (RD), both observables combine $\mathcal{S}$ and $\mathcal{A}$ modes in independent ways, so that $\hat I_\mathrm{cos}$ and $\hat I_\mathrm{sin}$ carry information on both modes~\cite{pra2013}. 

\textit{Phase mixing and the quantum state} -- Measurement of the photocurrent components $\hat I_{\cos{}}$ and $\hat I_{\sin{}}$ requires a well defined phase relation between LO and eLO, a condition in general \textit{not satisfied} in experiments. 
In fact, that limitation does not affect each individual quantum measurement of the spectral photocurrent, typically carried out on a time scale much shorter than the characteristic time of LO phase diffusion; they realize the pure quadrature observable 
\begin{equation}
\hat I_\theta=\cos\theta\,\hat I_\mathrm{cos}+\sin\theta\,\hat I_\mathrm{sin} \;.
\label{eq:imix}
\end{equation}
However, the collection of quantum statistics requires many such single measurements and hence a time interval that greatly surpasses the relative coherence time between LO and eLO. As a result, photocurrent \textit{moments} involve averaging different quadrature directions in phase space, entailing mixed quantum statistics ($\theta$-averages of moments of $\hat I_\theta$ on the quantum state). Let us construct the operator $\delta\hat I_\theta=\hat I_\theta-\ave{\hat I_\theta}$ with zero average on the quantum state, and consider its $\theta$-average variance, denoted as $\Delta^2\hat I_\theta$. Using Eq.~(\ref{eq:imix}), 
\begin{align}
\label{eq:Sruido}
\Delta^2\hat I_\theta &\equiv {\textstyle\frac{1}{2\pi}}\int d\theta'\ave{(\delta\hat I_{\theta+\theta'})^2}\nonumber\\
& = {\textstyle\frac{1}{2}}\Delta^2\hat I_{\cos{}}+{\textstyle\frac{1}{2}}\Delta^2\hat I_{\sin{}}, \;\forall\theta,
\end{align}
a quantity actually independent of the choice of $\theta$ (i.e. no LO-eLO relative phase information remains, as expected). Furthermore, components in quadrature ($\delta\hat I_\theta$ and $\delta\hat I_{\theta+\pi/2}$) are uncorrelated,  
\begin{align}
\ave{\delta\hat I_\theta\delta\hat I_{\theta+\pi/2}} &\equiv {\textstyle\frac{1}{2\pi}}\int d\theta'\ave{\delta\hat I_{\theta+\theta'}\delta\hat I_{\theta+\pi/2+\theta'}}\nonumber\\
& = \ave{\delta\hat I_{\cos{}}\delta\hat I_{\sin{}}-\delta\hat I_{\sin{}}\delta\hat I_{\cos{}}} =0, \;\forall\theta,
\label{eq:corrphasemix}
\end{align}
since $[\delta\hat I_\theta,\delta\hat I_{\theta+\pi/2}] = 0$. 
In this so-called \textit{phase mixing regime}, in which most experiments dealing with the spectral quantum noise of light are realized, Eqs.~(\ref{eq:Sruido}) and ~(\ref{eq:corrphasemix}) show that any electronic component of the spectral photocurrent displays the same quantum statistics and is uncorrelated with its simultaneously measured (in quadrature) component: in other words, the measured spectral photocurrent is stationary~\cite{pra2013}. But since stationarity holds simply as a consequence of phase mixing, it does not convey any information on the quantum state; in fact, all such information lies in the variance $\Delta^2\hat I_\theta$, the quantity proportional to the spectral noise power $S(\Omega)$ of semi-classical models, 
\begin{equation} 
\label{eq:Smixcossin}
S(\Omega) = \ave{\delta\hat I_\Omega\delta\hat I_{-\Omega}} =  {\textstyle\frac{1}{2}}\Delta^2\hat I_{\cos{}}+{\textstyle\frac{1}{2}}\Delta^2\hat I_{\sin{}}.
\end{equation}
That is probably the reason why in most experiments the spectral photocurrent is measured directly on a spectrum analyzer without further concern about the pure measurement operators $\hat I_{\cos{}}$ and $\hat I_{\sin{}}$, given that their statistical properties seem inaccessible~\cite{yurkeWidebandPRA85}. Nevertheless, the current approach to the measurement of spectral moments as solely based on the mixed moment of Eq.~(\ref{eq:Smixcossin}) can not be considered a pure quantum measurement for most quantum states~\cite{ralphSinglePhotonSidebandsPRA08,pra2013}. Furthermore, it fails to provide all the information required to reconstruct the two-mode spectral quantum state. Let us consider Gaussian quantum states (i.e. the simplest class of interesting quantum states) to illustrate this point. The complete reconstruction of any quantum state in this set would require the measurement of 10 independent second-order moments, gathered in the two-mode covariance matrix as 
\begin{equation}
\mathbf{V}=\left(
\begin{array}{cccc}
\Delta^2 \hat p_s & C(\hat p_s\hat q_s) & C(\hat p_s\hat p_a) & C(\hat p_s\hat q_a) \\
C(\hat p_s\hat q_s) & \Delta^2 \hat q_s & C(\hat p_a\hat q_s) & C(\hat q_s\hat q_a)\\
C(\hat p_s\hat p_a) & C(\hat p_a\hat q_s) & \Delta^2 \hat p_a & C(\hat p_a\hat q_a)\\
C(\hat p_s\hat q_a) & C(\hat q_s\hat q_a) & C(\hat p_a\hat q_a) & \Delta^2 \hat q_a
\end{array}
\right),
\label{eq:V2mode}
\end{equation}
where quadrature operator variances are denoted as in e.g. $\Delta^2 \hat p_s = \ave{(\hat p_s-\ave{\hat p_s})^2}$ and symmetrized correlations as in e.g. $C(\hat p_s\hat q_s) =\ave{(\hat p_s-\ave{\hat p_s})(\hat q_s-\ave{\hat q_s})+(\hat q_s-\ave{\hat q_s})(\hat p_s-\ave{\hat p_s})}/2$. It is clear that measurements of $S(\Omega)$ defined in Eq.~(\ref{eq:Smixcossin}) can provide only a fraction those 10 moments, although the exact amount of information is dependent on the particular measurement technique in use. 
HD can only retrieve 3 combinations of moments, while RD accesses 1 additional combination~\cite{pra2013}. However, if one could see beyond the phase mixing process to establish the phase-locked photocurrent components $\hat I_{\cos{}}$ and $\hat I_{\sin{}}$ as stationary, 6 combinations of moments could be quantified as null and the covariance matrix would simplify to  
\begin{equation}
\mathbf{V}=
\left(
\begin{array}{cccc}
\alpha & \gamma & \delta & 0\\
\gamma & \beta & 0 & \delta\\
\delta & 0 & \beta & -\gamma\\
0 & \delta & -\gamma & \alpha
\end{array}
\right). 
\label{eq:symcovmatrix}
\end{equation}
In this case, HD would be able to access the local single-mode sector of the quantum state (either $\mathcal{S}$ or $\mathcal{A}$), and RD would provide complete two-mode quantum state reconstruction (by additionally accessing the $\delta$ moment related to the energy asymmetry of spectral modes $\pm\Omega$)~\cite{prl2013}. 

Hence establishing the stationarity of the phase-locked photocurrent components $\hat I_{\cos{}}$ and $\hat I_{\sin{}}$ (or tacitly assuming it) stands as a central part of spectral quantum state measurement, either in the single-mode picture (HD) or in the complete two-mode picture (RD). We show next how it is possible to ascertain with great confidence the stationary property even in the presence of phase mixing, by analyzing \textit{higher order} moments of the spectral photocurrent. The reasoning we present provides the necessary conceptual basis to correctly interpret the photocurrent quantum noise in terms of moments of the symmetric covariance matrix of Eq.~(\ref{eq:symcovmatrix}), clarifying the tacit assumptions employed in nearly all experiments with spectral modes of light in the CV regime. 

\textit{Gaussian character of the measured photocurrent} -- We investigate next what can be said about the two-mode quantum state by having access only to the mixed operator moment of Eq.~(\ref{eq:Sruido}). Particularly, we ask what can be learned about the quantum state when the photocurrent is faithfully established by the experiment as Gaussian or non-Gaussian. Let us consider how higher order moments of the photocurrent relate to higher order quadrature operator moments in the phase mixing regime. 

We denote the (phase mixed) $2n$\textsuperscript{th}-order photocurrent moment as $\sigma^{\{2n\}}$, $n=1,2,3,\dots$, defined as 
\begin{equation}
\label{eq:mixedmoments}
\sigma^{\{2n\}}={\textstyle\frac{1}{2\pi}}\int_{-\pi}^{\pi}\ave{(\delta\hat I_\theta)^{2n}} d\theta.
\end{equation}
These measured moments can be expressed in terms of quadrature operator moments (and therefore related to the two-mode quantum state), through the measurement operators $\delta\hat I_\mathrm{cos}$ and $\delta\hat I_\mathrm{sin}$ of Eq.~(\ref{eq:imix}). These observables correspond to what would be measured with doubly phase-coherent detection, and represent the phase-locked photocurrent components masked by the phase mixing as well as pure independent measurements of the spectral quantum state. In order to relate those quantities to the mixed photocurrent moments of Eq.~(\ref{eq:mixedmoments}), we denote the moments of the phase-locked photocurrent as $\sigma_\mathrm{cos}^{\{2n\}}=\ave{(\delta\hat I_\mathrm{cos})^{2n}}$ and $\sigma_\mathrm{sin}^{\{2n\}}=\ave{(\delta\hat I_\mathrm{sin})^{2n}}$. 

For the particular but important case of Gaussian statistics (applicable to both the quantum state and the measured photocurrent), odd moments (other than the first) are null and all even moments can be expressed in terms of the variance of the distribution. This fact motivates us to denote the standard deviations of distributions as follows: $s = \sqrt{\sigma^{\{2\}}}$ for the measured (mixed) photocurrent, and $s_{\cos{}} =\sqrt{\sigma_\mathrm{cos}^{\{2\}}}$ or $s_{\sin{}} =\sqrt{\sigma_\mathrm{sin}^{\{2\}}}$ for the cosine or sine components subjacent to phase mixing. We also denote their normalized correlation (equal to the correlation between spectral modes) as $c\equiv\ave{\delta\hat I_\mathrm{cos}\delta\hat I_\mathrm{sin}}/(s_\mathrm{cos}s_\mathrm{sin})$ ($-1\leq c\leq1$). 

We would like to establish how the Gaussian character of the photocurrent constrains the field quantum state. Let us consider for  example the fourth-order moment $\sigma^{\{4\}}$ of the mixed photocurrent. On the one hand, a Gaussian photocurrent would imply $\sigma^{\{4\}}=3 s^4$; on the other, Gaussian quantum states would require fourth-order moments fulfilling $\sigma_\mathrm{cos}^{\{4\}} = 3s_\mathrm{cos}^4$, $\sigma_\mathrm{sin}^{\{4\}} = 3s_\mathrm{sin}^4$ and $\ave{\delta\hat I_\mathrm{cos}^2\delta\hat I_\mathrm{sin}^2} = (1+2c^2)s_\mathrm{cos}^2 s_\mathrm{sin}^2$.

To better analyze the actual distributions in terms of the Gaussian statistics, we define `deviations', i.e. quantities constructed to be null for Gaussian distributions, as follows: $\delta=\sigma^{\{4\}}-3s^4$ for the measured photocurrent and, for the quantum state, the quantities  $\delta_\mathrm{cos}=\sigma_\mathrm{cos}^{\{4\}}-3s^4_\mathrm{cos}$, $\delta_\mathrm{sin}=\sigma_\mathrm{sin}^{\{4\}}-3s^4_\mathrm{sin}$ and $\delta_c=\ave{\delta\hat I_\mathrm{cos}^2\delta\hat I_\mathrm{sin}^2} -(1+2c^2)s_\mathrm{cos}^2s_\mathrm{sin}^2$. 

We then employ Eqs.~(\ref{eq:imix}) and~(\ref{eq:mixedmoments}) to relate the (fourth-order) photocurrent deviations and quantum state deviations from the Gaussian statistics by 
\begin{equation}
{\textstyle\frac83}\delta = \delta_\mathrm{cos} +\delta_\mathrm{sin} +2\delta_c+\left(s^{2}_\mathrm{cos}-s^{2}_\mathrm{sin}\right)^2 +4c^2s_\mathrm{cos}^2s_\mathrm{sin}^2. 
\label{eq:fourthorder}
\end{equation} 
Interpretation of Eq.~(\ref{eq:fourthorder}) requires careful consideration of possible experimental scenarios. We note that actual measurements can only access the left side of the equality above with the goal of determining all the terms on the right: an impossible task if no \textit{a priori} assumptions are allowed. Let us analyze the exact content of those assumptions in the usual spectral noise detection. 

The first possibility is that the experimental data establishes the photocurrent statistics as Gaussian, imposing the constraint $\delta=0$, i.e. 
\begin{equation}
\delta_\mathrm{cos} +\delta_\mathrm{sin} +2\delta_c=-\left(s^{2}_\mathrm{cos}-s^{2}_\mathrm{sin}\right)^2 - 4c^2s_\mathrm{cos}^2s_\mathrm{sin}^2. 
\label{eq:fourthordergaussian}
\end{equation}
This equation establishes that infinitely many quantum states can in principle give rise to Gaussian spectral photocurrent. However, their fourth- and second-order moments must fulfill a very stringent relation. Hence most quantum states satisfying Eq.~(\ref{eq:fourthordergaussian}), even though not ruled out by the data alone, would be very unusual, obliging us to consider different scenarios to help interpret the data. 

In the first scenario, we note that the quantum dynamics capable of producing such stringent quantum states is unlikely to be taking place in most experiments in quantum optics or atomic physics. These type of experiments usually employ at least one bright light beam to participate in the quantum dynamics, in which cases a linearized or mean field approach in general supports the onset of Gaussian quantum states. The observation of Gaussian mixed photocurrent [Eq.~(\ref{eq:fourthordergaussian})] together with the \textit{assumption} of Gaussian quantum states, implies the new condition
\begin{equation}
\left(s^{2}_\mathrm{cos}-s^{2}_\mathrm{sin}\right)^2 + 4c^2s_\mathrm{cos}^2s_\mathrm{sin}^2 = 0,
\label{eq:fourthordergaussiangaussian}
\end{equation}
which can only be fulfilled if 
\begin{equation}
s_{\cos{}} = s_{\sin{}} \quad \mathrm{and}\quad c = 0,
\label{eq:fourthorderconsequence}
\end{equation}
meaning that the cosine and sine photocurrent components must present the same variance and be uncorrelated: a statement of \textit{stationarity}~\cite{pra2013}. 
Even though phase mixing always leads to stationary mixed photocurrent, verifying that its fourth-order moment is compatible with the Gaussian statistics makes it a better case to establish as stationary the phase-locked photocurrent related to pure quantum measurements. Thus checking higher order photocurrent moments allows us to remove some hindrances of the phase mixing process and partially `see through it'. 

As noted previously, the most relevant consequence of stationarity in our context can be established by interpreting Eq.~(\ref{eq:fourthorderconsequence}) in terms of properties of the spectral quantum state. Assuming it as Gaussian, stationarity implies the form of Eq.~(\ref{eq:symcovmatrix}) for the two-mode quantum state, which in turn allows us to employ only the mixed photocurrent variance as source of information about the quantum state.  
Only after this formal verification can one confirm that both HD and RD are able to access the \textit{single-mode} quantum state of either mode $\mathcal{S}$ or $\mathcal{A}$, justifying the application of entanglement criteria directly to the spectral noise matrix, equal in this case to the single-mode covariance matrix of the quantum state (semi-classical approach). If necessary, RD can be used to further access exclusive two-mode spectral features, a capability especially important in case one needs to reconstruct not only the local quantum state of $\mathcal{S}$ or $\mathcal{A}$ mode, but also the local quantum states of individual sideband modes $\pm\Omega$, in which case the energy imbalance becomes essential to perform the change of modal basis on the two-mode quantum state. 

We may summarize the first scenario that may be adopted to interpret the observation of \textit{Gaussian mixed photocurrent} as employing the following set of concepts in a self-consistent manner: Gaussian quantum state $\leftrightarrow$ Stationary phase-locked photocurrent $\leftrightarrow$ Symmetric $\mathcal{S}$ and $\mathcal{A}$ local quantum states. If one of these concepts is adopted as \textit{a priori} assumption to help interpret the experimental data (since none of them can be directly checked due to measurement mixedness) the others follow as consequences. For instance, in the case of phenomena associated with parametric downconversion, the broadband nature of the physical interaction will probably favor the adoption of `symmetric $\mathcal{S}$ and $\mathcal{A}$ local quantum states' as assumption, from which follow as consequences the validity of the semi-classical approach and the fact that an EPR-type quantum state is produced on the spectral sidebands (by a change of modal basis). 

In the second scenario, Eq.~(\ref{eq:fourthordergaussian}) allows the possibility that Gaussian photocurrent could be produced by non-Gaussian quantum states. Additional evidence would certainly be required in this case, since a Gaussian photocurrent would hardly have any credibility in attesting the non-Gaussian character of the quantum state. The type of quantum state capable of fulfilling Eq.~(\ref{eq:fourthordergaussian}) would have to present very unusual connections between fourth- and second-order moments. As we show later, the conditions to be fulfilled become more stringent as moments of even higher degree are considered, making it a difficult case for this kind of interpretation without further evidence. 

Since non-Gaussian features are desirable as a resource for quantum information protocols in CV, we consider the third scenario in which non-Gaussian mixed photocurrent could be used as misleading evidence of non-Gaussian features of the quantum state. In fact, according to Eq.~(\ref{eq:fourthordergaussian}), Gaussian quantum states impose upon the 4\textsuperscript{th}-order moment of the photocurrent the condition 
\begin{equation}
{\textstyle\frac83}\delta = \left(s^{2}_\mathrm{cos}-s^{2}_\mathrm{sin}\right)^2 +4c^2s_\mathrm{cos}^2s_\mathrm{sin}^2.
\label{eq:fourthordergaussiannongaussian}
\end{equation}
This identity states very generally that whenever the two individual $\mathcal{S}$ or $\mathcal{A}$ single-mode quadratures follow different Gaussian distributions, i.e. $\langle(\delta \hat I_{\cos{}})^2\rangle\neq\langle(\delta \hat I_{\sin{}})^2\rangle$, the photocurrent will present non-Gaussian statistics. The photocurrent deviation from Gaussian statistics could thus follow as a consequence of \textit{mixed Gaussian} processes, in fact the favored explanation in view of the ubiquity of Gaussian states in quantum optics. Such cases could be spotted by investigating the complete statistical distribution of measured quantum fluctuations, wherein Gaussian mixed models could be employed to estimate the single-mode probability distributions subjacent to the mixing of the measurement process. Failure to identify mixed Gaussian distributions would support further investigating the non-Gaussian character of the quantum state and probably require additional experimental capabilities. Conversely, Eq.~(\ref{eq:fourthordergaussiannongaussian}) makes the observation of Gaussian photocurrent as a special event, since its occurrence is a strong indication of Gaussian quantum states possessing modal quantum state symmetry. 

The conclusions reached for the 4\textsuperscript{th}-order moments are accentuated by extending our analysis to moments of even higher degrees. We find that the observation of Gaussian photocurrent in all orders either substantiates the Gaussian character of the quantum state or imposes increasingly stringent criteria upon the moments of the non-Gaussian quantum states capable of producing it. We define the deviation $\delta^{\{2n\}}$ of the $2n$\textsuperscript{th}-order moment from Gaussian statistics as 
\begin{equation}
\label{eq:highevenmoments}
\delta^{\{2n\}}=\sigma^{\{2n\}} - (2n-1)!!\,s^{2n}.
\end{equation}
Similarly, the deviation $\delta_{\{2n,2k\}}$ from Gaussian statistics regarding the quantum state is defined as 
\begin{equation}
\delta_{\{2n,2k\}}=\ave{\delta\hat I_\mathrm{cos}^{2n}\delta\hat I_\mathrm{sin}^{2k}} - \frac{(2(n-k))!(2k)!}{2^n(n-k)!k!}\,s_{\cos{}}^{2(n-k)}s_{\sin{}}^{2k},
\end{equation}
where $k=0,1,2,\dots,n$ and $s = s_{\cos{}} = s_{\sin{}}$. Performing the integral in $\theta$, Eq.~(\ref{eq:mixedmoments}) applied to higher-order moments yields
\begin{equation}
\sigma^{\{2n\}}=\frac{n!}{(2n)!}\sum_{k=0}^n\frac{1}{(n-k)!k!}\ave{(\delta\hat I_{\cos{}})^{2n}(\delta\hat I_{sin{}})^{2(n-k)}}.
\end{equation}
In the following we suppose uncorrelated single modes (i.e. $c=0$), a simplifying restriction that does not affect the physical conclusions. 
The connection between photocurrent statistics and quantum state statistics provided by any even moment of order larger than 2 is 
\begin{align}
\label{eq:constrainthigher}
 \delta^{\{2n\}} - \frac{n!}{(2n)!}\sum_k &\frac{1}{(n-k)!k!}\, \delta_{\{2(n-k),2k\}} = \\
&= \frac{(2n-1)!!}{2^n}\sum_k d_{n,k}s_{\cos{}}^{2(n-k)}s_{\sin{}}^{2k},\nonumber
\end{align}
where the coefficients are $d_{n,k}= \frac{n! - (2(n-k)-1)!!(2k-1)!!}{(n-k)!k!}$. The left-hand side of this expression relates only to \textit{photocurrent} deviations from the Gaussian statistics, and the right-hand side only to \textit{quantum state} deviations. In fact, $\delta^{\{2n\}}$ is the only parameter in Eq.~(\ref{eq:constrainthigher}) available in the data. The quantum state must be determined from incomplete information due to the mixing process affecting the measurement operator. This is indeed an impossible task if any quantum state can be expected in the experimental dynamics under study. However, as noted before, usually that is not the case, and in fact the quantum dynamics producing non-Gaussian states constrained by Eq.~(\ref{eq:constrainthigher}) for a given measured value of $\delta^{\{2n\}}$ is very difficult to attain in the laboratory. Such state of affairs calls for a detailed analysis of Eq.~(\ref{eq:constrainthigher}) in particular cases of interest.

In case the photocurrent is Gaussian, i.e. $\delta^{\{2n\}}=0, \forall n$, Eq.~(\ref{eq:constrainthigher}) provides very stringent restrictions connecting the quantum state deviations from the Gaussian statistics with second-order operator moments. Those types of quantum states can usually be ruled out based on physical considerations of the quantum dynamics acting on the system. It is thus more plausible in this case to interpret the Gaussian photocurrent as indicating the Gaussian character of the quantum state, for which the left-hand side of Eq.~(\ref{eq:constrainthigher}) must vanish (i.e. $\delta_{2(n-k),2k} =0, \forall n,k$). Assuming then the quantum state as Gaussian, one finds that the right-hand side of Eq.~(\ref{eq:constrainthigher}) can only vanish (as required to uphold the identity) if $s_{\cos{}} = s_{\sin{}}$, in which case it is simple to verify that the sum of coefficients yields $\sum_k d_{n,k} = 0$. We find once more the self-consistent connections between Gaussian quantum states, symmetric $\mathcal{A}$ and $\mathcal{S}$ modes, and stationarity of the phase-locked photocurrents subjacent to the mixing process. Conversely, a Gaussian quantum state will produce non-Gaussian photocurrent with respect to every higher-order moment if $s_{\cos{}} \neq s_{\sin{}}$. 

It is a striking fact that Gaussian photocurrent together with the assumption of Gaussian quantum state will always lead to the equality of the single modes $\mathcal{A}$ and $\mathcal{S}$ quadrature distributions, and vice-versa. 
Similarly simple conclusions can not be reached for non-Gaussian photocurrents, since they could result either from truly non-Gaussian quantum states or from a mixture of two different Gaussian processes (i.e. when both single-mode quadrature distributions of mode $\mathcal{S}$ and $\mathcal{A}$ are Gaussian, but different).

\section{Experiment and data analysis}
\label{sec:experiment}

\textit{Experimental setup} -- The experiment (Fig.\ref{fig:setup}) employs a frequency-doubled diode-pumped Nd:YAG cw laser (Innolight Diabolo) at 532 nm wavelength to pump a nondegenerate triply resonant optical parametric oscillator (OPO). Prior to injection in the OPO, the cw pump laser is filtered in transmission by an optical resonator (bandwidth of 0.7 MHz) to attenuate classical noise for analysis frequencies above 15 MHz. In this manner, the pump laser shows nearly vacuum noise limited quantum fluctuations at the spectral modes of interest at $\Omega = 21$~MHz. The OPO consists of a type-II phase matched KTP (Potassium Titanyl Phosphate, KTiOPO4) crystal with length $l=12$~mm and a linear resonator with spherical mirrors with 50~mm radius of curvature. The input coupler mirror has reflectivity of 70\% for the pump (532 nm) and $>99.9\%$ for the signal and idler beams ($\approx$~1064 nm), and the output coupler mirror has transmission of 96\% for the infrared beams and is highly reflective for the pump beam. The OPO has a free spectral range of about 5 GHz and cavity finesses of 18, 135, and 115 for pump, signal and idler modes, respectively. The KTP crystal temperature is actively stabilized at 23\textsuperscript{o}C. The OPO oscillation threshold power is $P_{th}=60(3)$~mW. Operating above the threshold, it generates collinear signal and idler beams with orthogonal polarizations, which are then spatially separated by a polarizing beam-splitter (PBS). The reflected pump beam is separated from the input beam by a circulator employing a Faraday rotator (RF) and a PBS. 

We employ the RD technique to reconstruct the quantum states of spectral modes in the pump, signal and idler beams individually~\cite{prl2013}. In addition to being a complete quantum measurement for Gaussian quantum states, in our experiment RD has also the advantage of allowing the use of the bright beams produced by the OPO as independent LO's at the appropriate optical frequencies. One analysis resonator is employed to perform RD in each beam. Spatial mode matching between the beams and the analysis resonators is higher than 95\%. The analysis resonator labeled $j=0$, corresponding to the pump mode, has spectral bandwidth of $\approx$~12 MHz; resonators labeled $j=1$ and $j=2$, respectively used to measure signal and idler beams, have bandwidths of $\approx$~14 MHz. With these values it is possible to access all quadratures for the chosen sideband modes at the chosen analysis frequency of 21 MHz. 

Photodetection is performed with high quantum efficiency ($>95\%$) photodiodes (Epitaxx ETX300 for the twin beams and Hamamatsu S5973-02 for the pump beam). Quantum noise of each beam is measured with a pair of amplified photodetectors using the balanced detection scheme. The Standard Quantum Level (SQL) of the shot noise is obtained by subtracting their photocurrents. The overall detection efficiencies accounting for photodetector efficiencies and losses in the optical paths are 87\% for the twin beams and 65\% for the pump beam. Electronic noise is subtracted according to independent calibration of its Gaussian distribution. 

The photocurrent is spectrally analyzed by a demodulating module, where the temporal signal is electronically mixed with a sinusoidal reference (eLO) at $\Omega=21$~MHz, in this manner defining the frequency of spectral field modes of interest. The result of electronic mixing is filtered by a 600 kHz bandwidth low-pass filter and recorded by a computer with an analog-to-digital board (NI PCI-6110). For each combination of field quadratures on a single beam, 1,000 spectral quantum measurements are realized in order to obtain photocurrent moments. In total, 450 different directions in the two-mode phase space of each beam are probed with RD, providing a total of 450,000 spectral quantum measurements per beam in 750 ms acquisition time. 

\begin{figure}[htbp]
\caption{Experimental setup. Nd:YAG Diabolo: Pump laser; OPO: Optical parametric oscillator;
PBS: Polarizing beam splitter; FR: Faraday rotator; KTP: Nonlinear crystal.}
\label{fig:setup}
\end{figure}

\textit{Experimental results} -- We now examine the quantum statistics of photocurrent fluctuations of the tripartite system composed of the pump, signal, and idler light beams produced by the above-threshold OPO. Our aim is to provide a detailed study of the Gaussian or non-Gaussian character of the resulting quantum states, which involve in principle 6 spectral modes (a pair for each beam)~\cite{prl2013}. Different quantum state can be produced by controlling the input pump power $P=P_0/P_{th}$ (measured relative to the threshold power). For each of them, we probe with RD the quadrature probability distribution in a given direction in the phase space of quadrature observables. 

\begin{figure}[htbp]
\includegraphics[width=0.49\columnwidth]{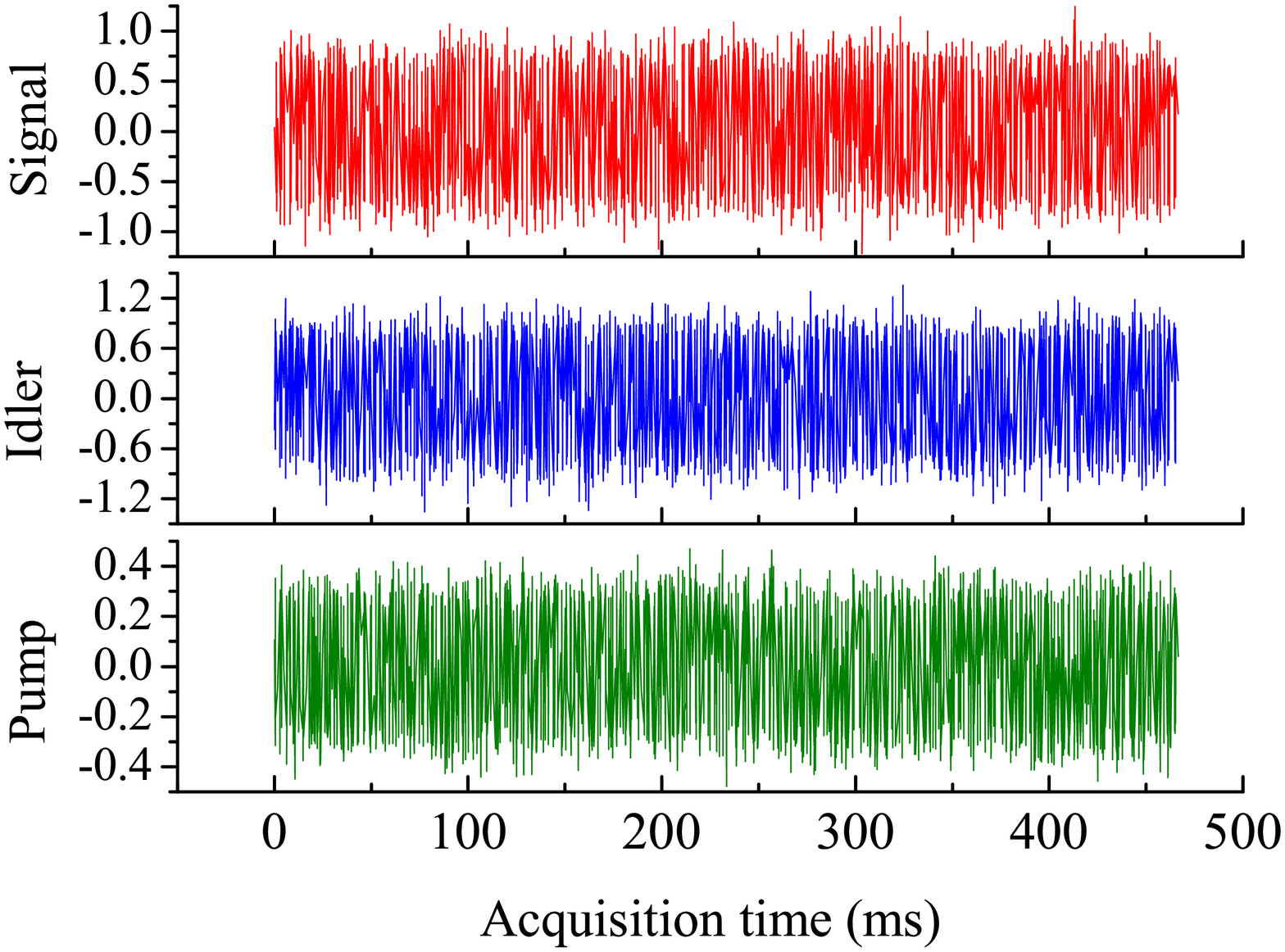}
\includegraphics[width=0.49\columnwidth]{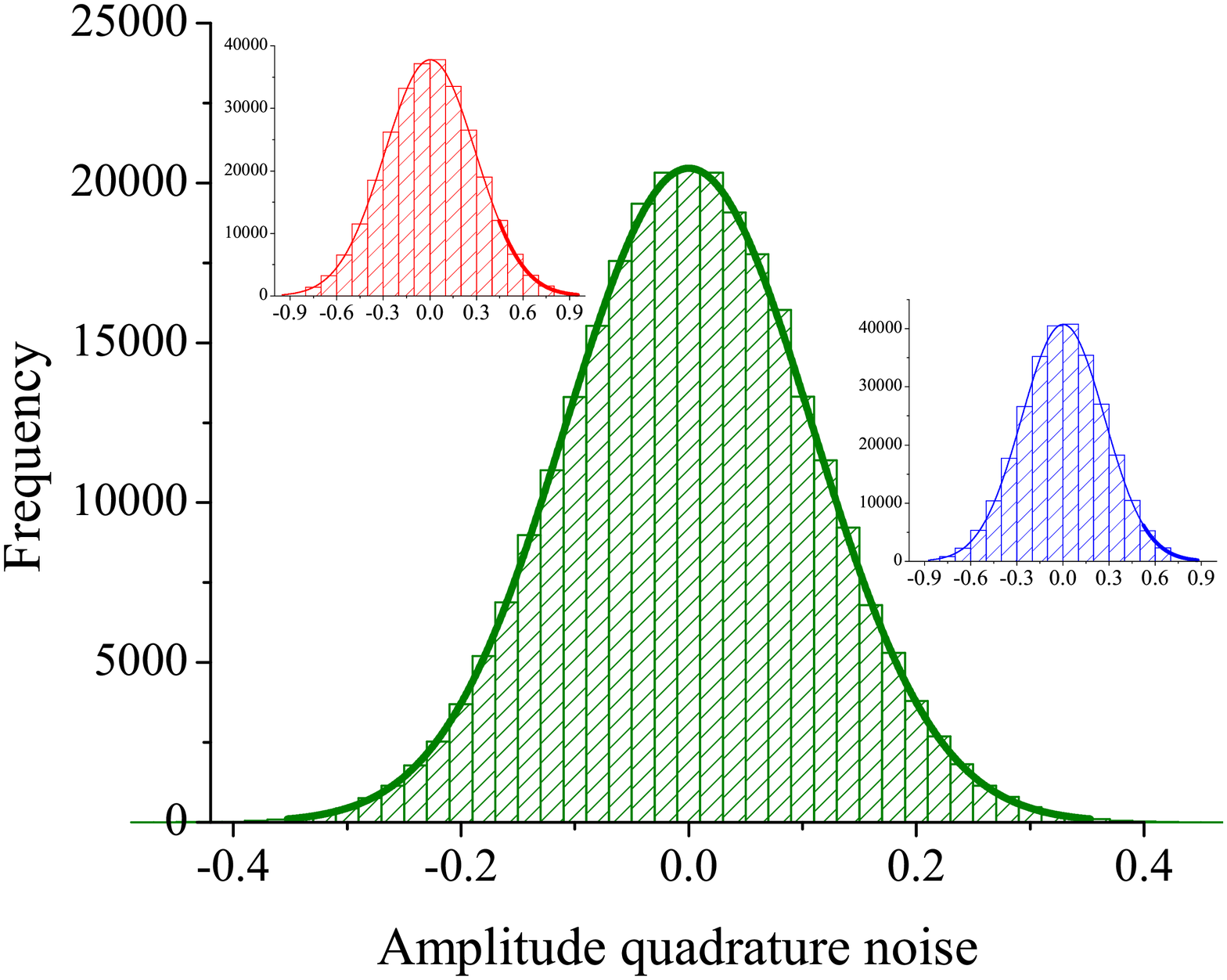}
\caption{(a) Raw data of quadrature amplitude fluctuations obtained with the analysis resonator far off resonance. (b) Histograms of raw data with the normal density distribution superimposed.}
\label{fig:histo}
\end{figure}

We start by analyzing the quantum states of each individual beam (2 spectral modes) for different values of pump power $P$. For each quantum state (i.e. each value of pump power), we perform a thorough investigation of a single quadrature direction in phase space. Focusing on a single quadrature observable allows us to gather larger statistical samples and analyze the Gaussian character of the photocurrent up to the 14\textsuperscript{th}-order moment ($n=7$). 
The chosen quadrature corresponds to the semi-classical amplitude quadrature and is measured with the analysis resonators far off resonance ($\Delta\gg1$). We collect $N=$~280,000 consecutive quadrature quantum measurements for each quantum state. One such data set, for $P=1.72$, is presented in Fig.~\ref{fig:histo}. From top to bottom on the left column, idler, pump, and signal spectral photocurrent fluctuations are shown. The corresponding histograms are presented on the right column of Fig.~\ref{fig:histo}. They show clear visual agreement with the normal distribution. 

\begin{figure}[htbp]
\begin{flushright}
\includegraphics[width=0.98\columnwidth]{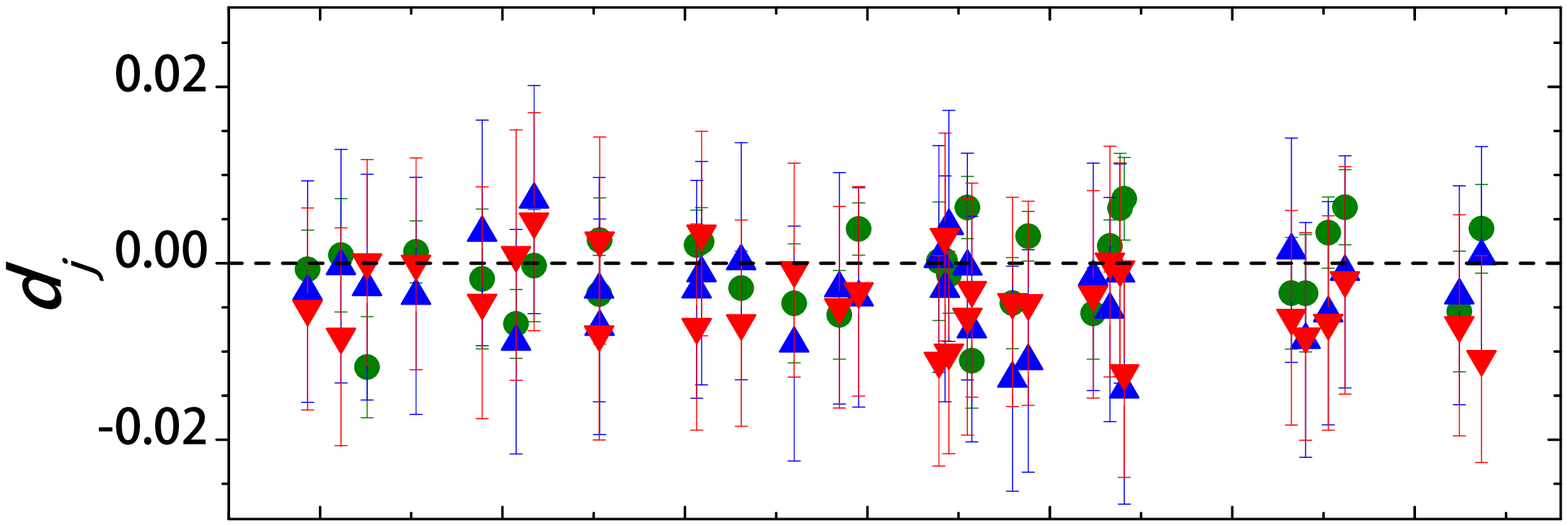}
\includegraphics[width=0.98\columnwidth]{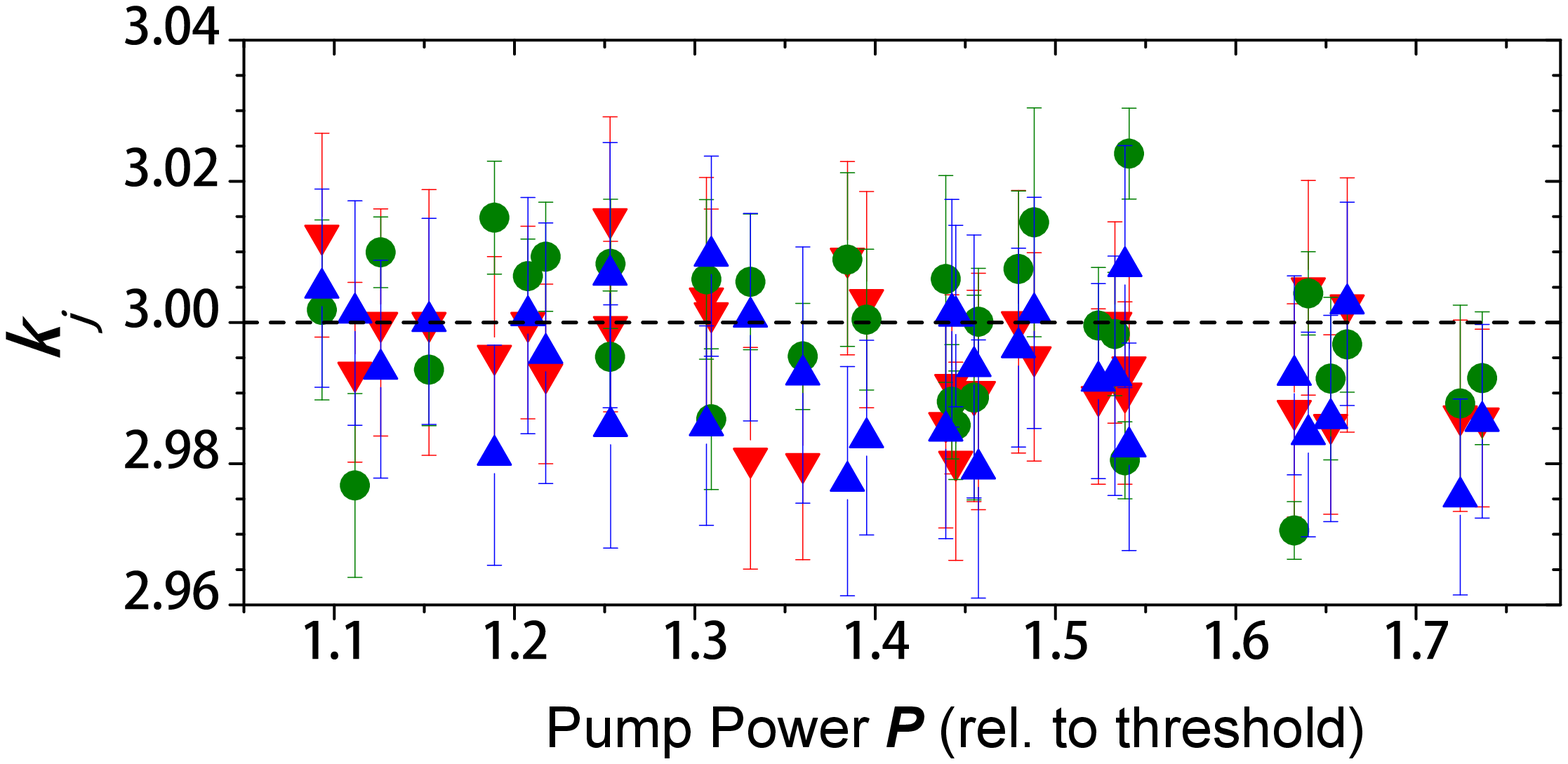}
\end{flushright}
\caption{Statistical analysis of 3\textsuperscript{rd}- (top) and 4\textsuperscript{th}-order (bottom) moments of single-beam photocurrent measurements relative to the distribution standard deviation. Dashed lines represent the values expected for  Gaussian statistics. Green circle: pump; Blue up-pointing triangle: signal; Red down-pointing triangle: idler. }
\label{fig:moments34}
\end{figure}

Quantitative analysis of the Gaussian character of those signals is first performed by considering the 3\textsuperscript{rd}- and 4\textsuperscript{th}-order moments, shown in Fig.~\ref{fig:moments34}. We define the third order moment $d_j=\sigma_j^{\{3\}}/s_j^3$, related to the skewness of the probability distributions, and fourth-order moment $k_j=\sigma_j^{\{4\}}/s_j^4$, akin to the kurtosis. Indices $j=0,1,2$ indicate pump, signal, and idler beams, respectively. For each value of pump power, moments are extracted from the complete set of 280,000 data points. For Gaussian statistics, one expects $d_j=0$ and $k_j=3$, as indicated by the black dashed lines on the figures. 

According to the results in Fig.~\ref{fig:moments34}, all measured fourth-order moments are compatible with the Gaussian distribution, meaning that the photocurrent signal of any single beam produced by the above-threshold OPO would fulfill Eq.~(\ref{eq:fourthordergaussian}) within the experimental uncertainty. In the case of $d_j$, the null result does not provide information about the quantum state, since it is a consequence of the phase mixing regime [Eq.~(\ref{eq:mixedmoments})]. For both quantities the fluctuations observed in the datasets are compatible with the statistical uncertainty of each data point. The number $N$ of quantum measurements per data trace alone would allows us to establish the Gaussian character of the photocurrent within $1/\sqrt{N}\approx~$0.2\% of the moment values. However, at these levels of confidence, we could notice by the observation of higher than expected discrepancy between different experimental data points the presence of a systematic error affecting the mean value of the photocurrent. This error source, probably caused by the photodetection electronics and still under investigation, has been included on the overall uncertainty of our data. The overall values and uncertainty for pump, signal, and idler fourth-order moments considering the whole data set as one are respectively $k_0 = 2.9987 (17)$, $k_1 = 2.9951 (26)$, and $k_2 = 2.9931 (27)$. The third-order moments are $d_0 = -0.0006 (9)$, $d_1 = -0.0041 (21)$, and $d_2 = -0.0033 (23)$. 

\begin{figure}[htbp]
\includegraphics[width=0.98\columnwidth]{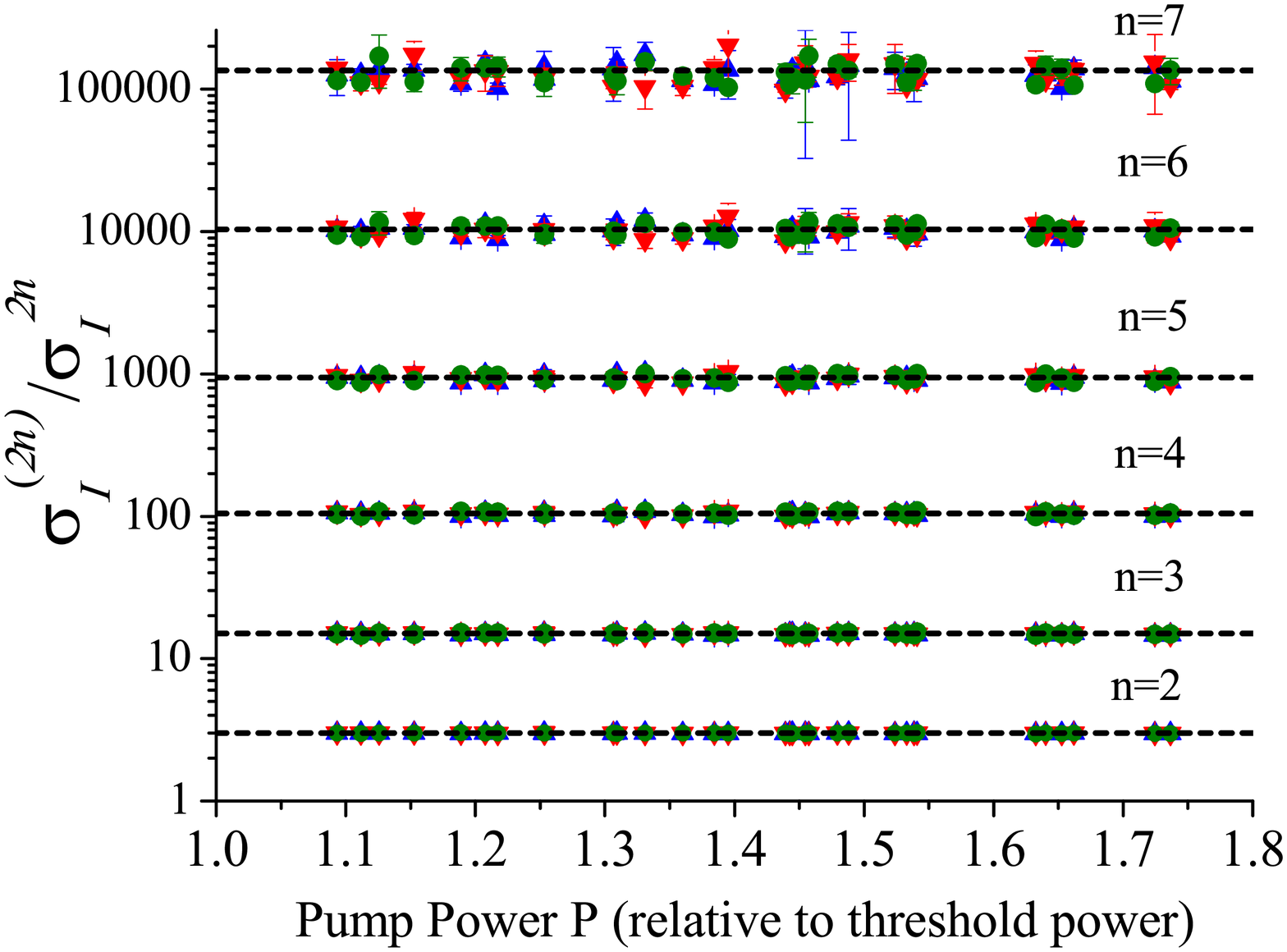}
\caption{Statistical analysis of higher-order moments of single-beam photocurrent measurements relative to the distribution standard deviation. Dashed lines represent the values expected for  Gaussian statistics. Green circle: pump; Blue up-pointing triangle: signal; Red down-pointing triangle: idler.  }
\label{fig:moments14}
\end{figure}

Moments of higher order are now probed in the same way, by keeping the same choice of quadrature measurement ($\Delta\gg1$). Results are presented in Fig.~\ref{fig:moments14}. We consider the ratios $r^{\{2n\}}= \sigma_j^{\{2n\}}/s_j^{2n}$, for $n=2,3,4,5,6,7$, where $s_j$ is the standard deviation of the quadrature probability distribution for beam $j$. Values expected for a Gaussian distribution are indicated by the dashed lines. All moments show perfect compatibility with the Gaussian distribution within the experimental uncertainty.

\begin{figure}[htbp]
\centerline{\includegraphics[width=1.08\columnwidth]{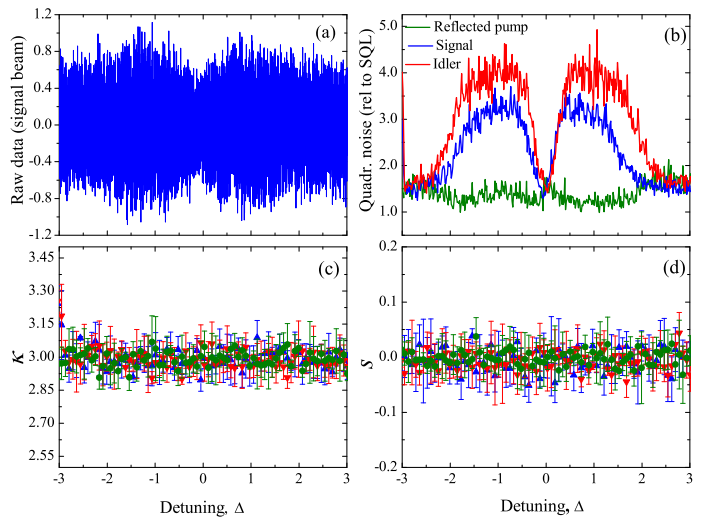}}
\caption{Statistical analysis as a function of resonator detuning. (a) Quantum fluctuations of the pump beam normalized by the average intensity (raw data). (b) Spectral quantum noise produced by pump (green curve, bottom), signal (blue curve, middle) and idler beams (red curve, top). (c) Fourth-order moments normalized by the standard deviation. (d) Third-order moment normalized by the standard deviation. Dashed lines represent the exact value expected from Gaussian statistics. In (c) and (d), symbols follow the same code of 
Fig.~\ref{fig:moments14}. }
\label{fig:cavdet}
\end{figure}

Next we investigate whether different directions in phase space could show deviations from the Gaussian statistics, by varying resonator detunings. We consider the quantum state produced by a fixed pump power $P$ = 1.66(5) and perform quadrature measurements with RD (similar results are obtained for different values of $P$). 
Fig.~\ref{fig:cavdet} depicts RD the spectral photocurrent of the three beams measured as functions of resonator detunings. Formally, each value of $\Delta$ corresponds to a direction of observation in the two-mode phase space. Inset (a) shows the measured quantum fluctuations for the pump beam as functions of $\Delta$ and normalized by its average intensity (raw data), corresponding to single measurements of the observable $\hat I_\theta(\Delta)$. Inset (b) presents the quantum noise (variance of photocurrent fluctuations) for the three single beams, relative to the SQL. Each point in the graph is a realization of Eq.~(\ref{eq:Sruido}), averaged over 1,000 single quantum measurements. While the pump beam presents nearly shot-noise limited quantum noise for all values of $\Delta$, signal and idler beams show excess noise in some directions in phase space (in the semi-classical picture, the beams would be described as possessing phase noise), determined by a theoretical curve used to obtain the optimum values of mixed operator moments. Fig.~\ref{fig:cavdet} (c) presents the results of $k_j$ for different quadratures of each single-beam, obtained by complete scans in $\Delta$. The statistical uncertainty in this case is around $1/\sqrt{1000}\approx $~3\%, and no significant deviation from the Gaussian statistics is found. Inset (d) displays once more a null value of $d_j$, a consequence of the phase mixing regime.

\begin{figure}[htbp]
\includegraphics[width=0.98\columnwidth]{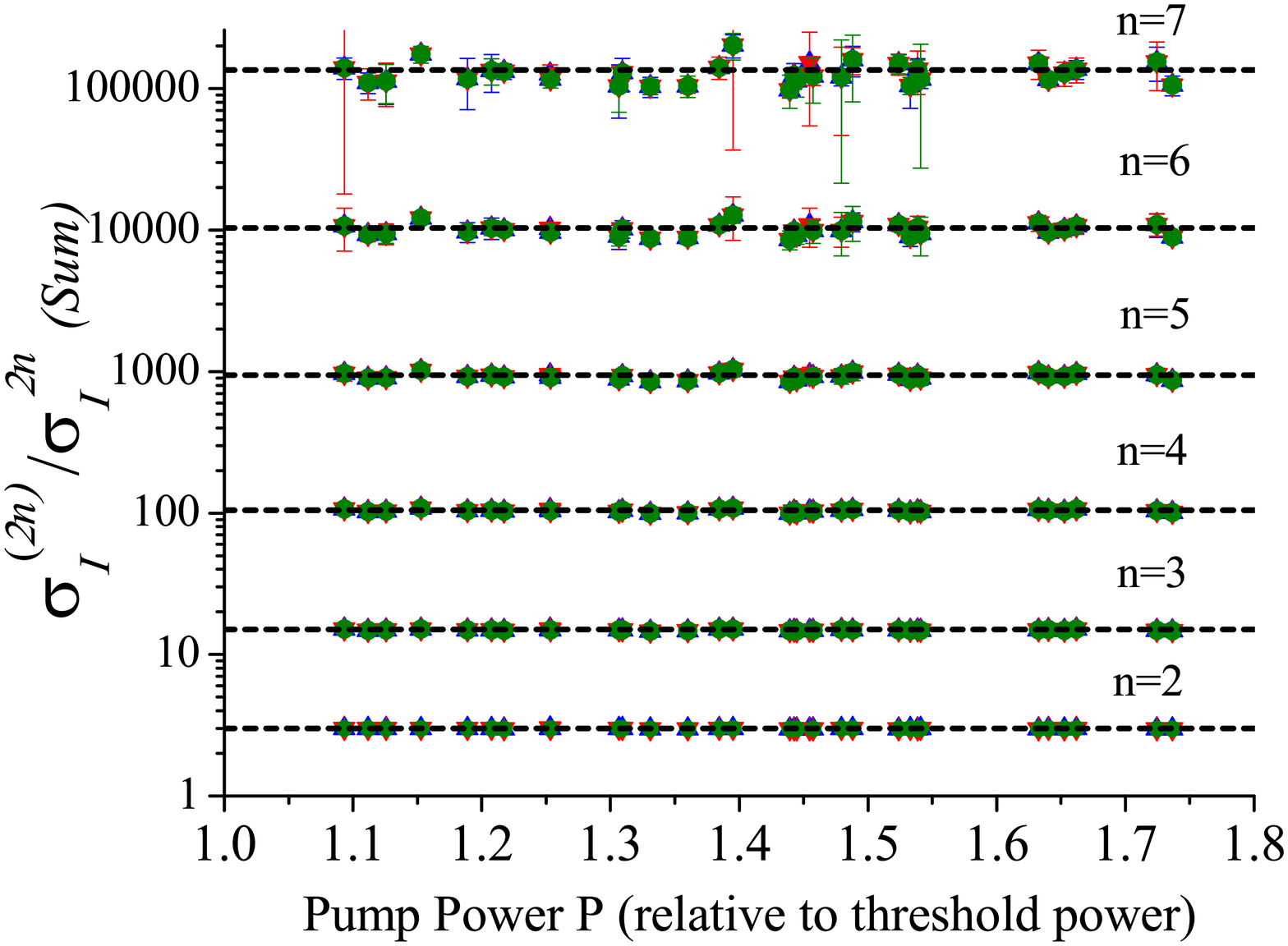}
\includegraphics[width=0.98\columnwidth]{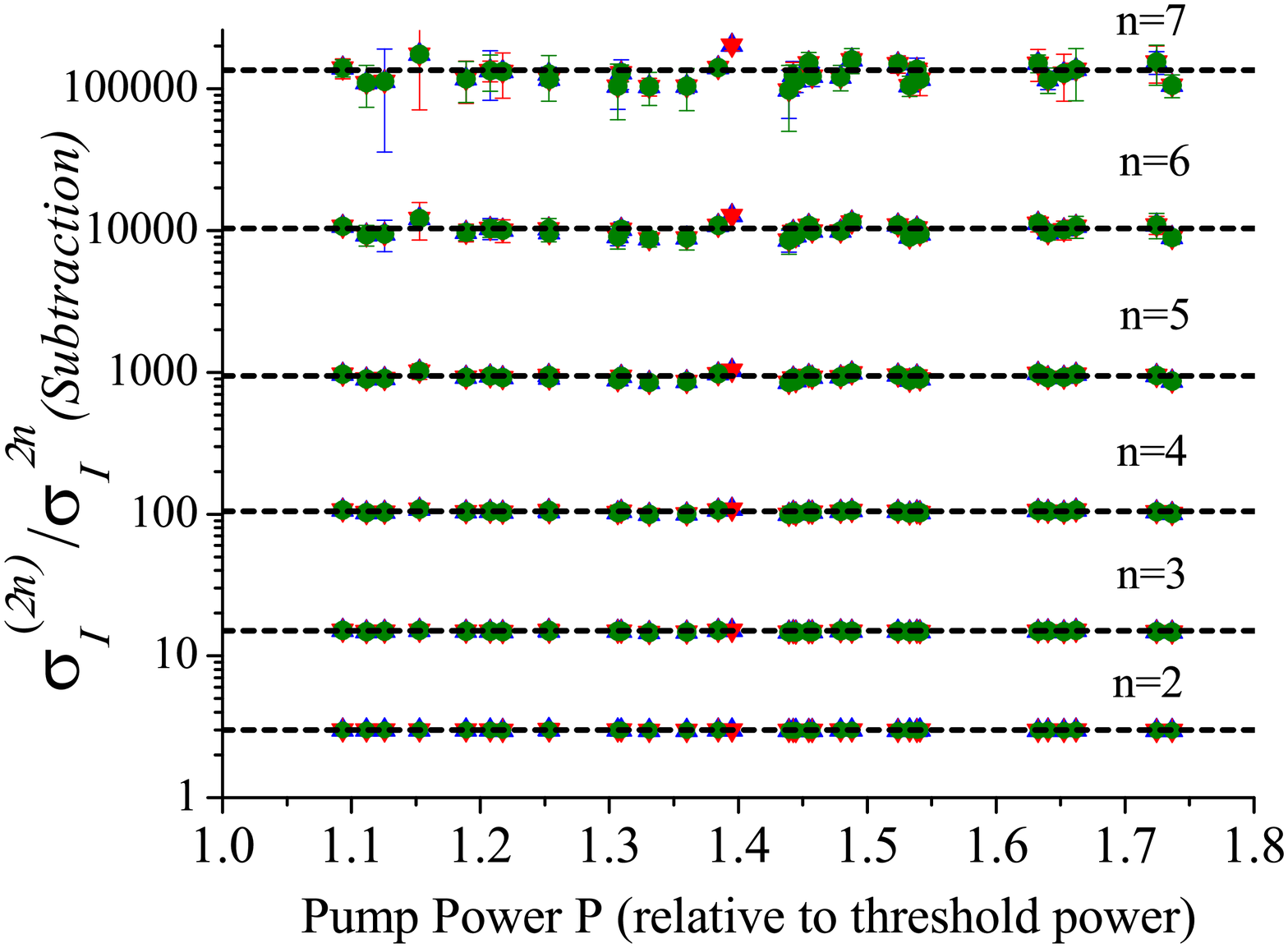}
\caption{Statistical analysis of higher order moments of two-beam photocurrent measurements normalized by the standard deviation of the respective probability distribution. Dashed lines represent the exact value expected from Gaussian statistics. Sum (top) and subtraction (bottom) of single-beam photocurrents are shown. Green circle: pump--idler combination; Blue up-pointing triangle: pump--signal combination; Red down-pointing triangle: signal--idler combination.}
\label{fig:momentssubsum}
\end{figure}

Finally, the analysis of two-beam correlations, in which four spectral modes are involved, is realized by changing modal basis to the sum or subtraction of single beam spaces. We consider the operators $\hat I^{(\pm,j,j')}=(\hat I^{(j)}_{\theta}\pm\hat I^{(j')}_{\theta})/\sqrt2$, $j\neq j'$, constructed by the coherent combination of single-beam observables. In this manner, two-beam probability distributions are probed with the tools employed above for single-beam quadratures. In fact, the presence of non-Gaussian correlations among beams would appear in this picture as non-Gaussian probability distributions of single quadrature observables related to two beams, by a change of modal basis. Fig.~\ref{fig:momentssubsum} summarizes the experimental results. No deviations from the Gaussian statistics can be observed for either the sum or the subtraction of single-beam photocurrents in the data. 

We have also applied the Shapiro-Wilk normality test, or W test, to further confirm the Gaussian character of the photocurrent signals. The test determines the degree of reliability by which the data sample would be incongruent with Gaussian statistics (the null-hypothesis). Shapiro and Wilk proved the power of their test showing its capability of detecting non-normality in small sets for a wide variety of statistical distributions, including those with Gaussian kurtosis values~\cite{shapiro1}. The test can be extended to sets containing up to thousands of samples. We have performed numerical investigations and found no advantage in the W test applied to our data, for which approximately 1\% accuracy is reached. Due to our large sets of data (hundreds of thousands), the analysis of higher order moments presented above reaches better precision.

\section{Conclusion} 
\label{sec:conclusion}

When employing the spectral analysis of the photocurrent to retrieve properties 
of the quantum state of light, it is important to realize that the Gaussian or 
non-Gaussian character of the observed photocurrent noise does not establish 
\textit{by itself} the type of statistics followed by the quantum state, since 
the photocurrent formally results from an incoherent mixture of two independent 
quantum measurements. Pure photocurrent operators, better understood in the modal 
basis of symmetric ($\mathcal{S}$) and antisymmetric ($\mathcal{A}$) combinations 
of spectral sideband modes, can only be retrieved if both the \textit{optical and 
electronic} local oscillators are phase coherent to one another during the whole process 
of quantum state reconstruction, in an implementation of a \textit{doubly phase-coherent} detection. 

Thus the observation of Gaussian statistics of the spectral photocurrent does not 
provide an unambiguous account of the quantum state statistics with the usual 
incoherent detection. It either strongly constrains the types of non-Gaussian quantum 
states capable of producing the Gaussian photocurrent or provides strong evidence 
for Gaussian and symmetric single-mode quantum states of $\mathcal{S}$ and 
$\mathcal{A}$ modes. A similar scenario holds in case non-Gaussian photocurrent 
fluctuations are observed. While it is true that non-Gaussian quantum states could 
generate the observed statistics, an incoherent mixture of Gaussian processes may also 
produce non-Gaussian quantum noise. 

The use of the photocurrent statistics to reconstruct the quantum state in general 
requires \textit{a priori} knowledge. In most experiments in quantum optics, at least 
one optical field with `classical' characteristics participates in the quantum dynamics, 
in which case a linearized interaction will favor the interpretation of data as 
stemming from \textit{Gaussian} and \textit{symmetric} quantum states in the $\mathcal{S}$/$\mathcal{A}$ 
modal basis (if one of 
those characteristics is taken as assumption, the other follows as consequence). In 
this scenario, the Gaussian character of the photocurrent implies that the phase-locked 
photocurrent components (i.e. the ones which would be measured with doubly phase-coherent 
detection) are themselves stationary. Thus, even though phase mixing does not allow direct 
verification of this property, the Gaussian character of the mixed signal indirectly 
establishes it, as long as the quantum state is assumed to be Gaussian (or, equivalently, 
the possible non-Gaussian quantum states that would lead to the same Gaussian photocurrent 
are dismissed as implausible). 

The correct reconstruction of the quantum state heavily 
relies on this fact, since only then can the covariance matrix be written in the simple 
form of an \textit{effective single-mode} Gaussian quantum state~\cite{pra2013}. Most experimental work in 
quantum optics employing spectral noise analysis assumes stationarity to hold without 
explicitly mentioning it, a limitation that could be a problem for the realization of 
CV quantum information tasks on those systems. Another consequence of the fact that 
most experiments with bright light beams will favor the generation of Gaussian states 
of the field lies in the interpretation of non-Gaussian photocurrent, which should first 
be likely attributed to a mixture of two different Gaussian distributions, one pertaining 
to the pure quadrature distribution of mode $\mathcal{S}$, and another to mode $\mathcal{A}$. 
To further substantiate the non-Gaussian character of the quantum state, the mixing of 
two Gaussian processes must first be ruled out, for instance by showing that mixed Gaussian 
models fail to account for the data. However, the unquestionable establishment of non-Gaussian 
properties of the spectral quantum state would require stronger evidence to be convincing, 
owing to its prized value as a resource in the realization of quantum information protocols. 

We apply these considerations to perform a thorough experimental investigation of the 
photocurrent noise from pump, signal, and idler beams produced by the above-threshold OPO. 
Our data establishes the photocurrent quantum noise of those beams as Gaussian within 
a broad range of investigated parameters. We analyze moments up to the 14\textsuperscript{th} 
order for individual beams and for cross-correlations. 
From a theoretical point of view, the OPO dynamics above the oscillation threshold is 
expected to be completely described by a linearized model (since only bright beams drive 
the interaction), from which Gaussian states of the field follow. The experimental 
observation of Gaussian photocurrent together with the well founded theoretical assumption 
of Gaussian quantum states implies the symmetry of $\mathcal{S}$ and $\mathcal{A}$ local 
quantum states for each beam. Our results demonstrate the stationarity of the phase-coherent 
spectral photocurrent to justify the use of the semi-classical model of quantum noise and 
thus validate our previous analysis of multipartite entanglement in this physical 
system~\cite{villarPRL05,science09,natphoton10}. Even though the complete state of the 
three beams entails six modes for a given analysis frequency, it is legitimate to restrict 
the analysis to three `effective' modes (choosing either set of the equivalent $\mathcal{S}$ 
or $\mathcal{A}$ modes), by a partial trace operation. The missing correlations between 
$\mathcal{S}$ and $\mathcal{A}$ modes, a feature yet to be measured, can only increase the 
strength of quantum correlations in case all six spectral modes are considered. 

We conclude that in most experiments \textit{a priori} knowledge is unavoidable to 
perform spectral quantum state reconstruction. A few exceptions are experiments which 
use both the LO and the eLO to generate the quantum state, as in Refs.~\cite{mlynekNature97,prl2013}. 
Such an improved situation can be applied to all experiments in quantum optics dealing 
with spectral modes by phase-locking the optical and electronic local oscillators, e.g. by 
employing sub-Hz linewidth lasers~\cite{subhertzlaser}, to realize doubly 
phase-coherent detection. This 
would ensure the phase coherence of all oscillators employed to extract the spectral quantum
 noise, making the quantum measurement associated with the spectral photocurrent a pure 
observable. We view such experimental improvement as essential to achieve assumption-free 
realization of quantum information protocols with CV spectral modes, particularly in the case of
quantum protocols requiring measurements of pure observables to be used as feedback 
to control the quantum state.

\acknowledgments

We acknowledge the support by  \#2010/52282- 1, \#2010/08448-2, S\~ao Paulo Research Foundation (FAPESP) projects, CNPq, and CAPES (through the PROCAD program). This research was performed within the framework of the Brazilian National Institute for Science and Technology in Quantum Information (INCT- IQ).

\end{document}